\documentstyle[prl,aps,preprint]{revtex}
\begin{document}
\textwidth6.8in
\textheight8.5in
\oddsidemargin -0.1in \evensidemargin -0.1in

\title{Relativistic $J$-matrix method}
\author{Pawe\l{} Horodecki \\
Faculty of Applied Physics and Mathematics\\
Technical University of Gda\'nsk, 80--952 Gda\'nsk, Poland}

\maketitle
\begin{abstract}
The relativistic version of the J-matrix method for
a scattering problem on the potential vanishing faster
than the Coulomb one is formulated.
As in the non-relativistic case it leads to
a finite algebraic eigenvalue problem.
The derived expression for the tangent of phase shift
is simply related to the non-relativistic case formula
and gives the latter as a limit case.
It is due to the fact that the used
basis set satisfies the ``kinetic balance condition''.
\end{abstract}
\vfill
\eject
%
%
%
\section{Introduction}
\noindent
The J-matrix method, introduced by Heller and Yamani \cite{Hell74a,Hell74b}
and developed by Yamani and Fishman \cite{matem}, is
an example of an algebraic method in
quantum scattering theory.
Comparing with
the algebraic variational theories the method
has been shown to be free of the false resonances problem
\cite{Hell}.
It has been used in construction of the
Gauss quadrature of the continuum \cite{Broad} (see also \cite{YR}),
the definition and analysis of a reproducing kernel in
the context of Harris eigenvalues \cite{Yamani}.
Quite recently, it has been used in formulation
of complex-scaling method  \cite{YA1}
and in development of the multi-channel Green's
functions \cite{YA2} by means of a complete $L^{2}$ basis.

The crux of the method
is representation of the Hamiltonian in a suitable non-orthogonal
basis changing the differential scattering problem
into the purely algebraic one.
Thus far only the non-relativistic version of the
method has been formulated. The aim of this paper
is to develop the simple relativistic formulation of the method
in its theoretical framework for potentials sufficiently
regular at the origin and vanishing at
infinity faster than the Coulomb one.
\section{Non-relativistic J-matrix - radial kinetic energy case}

First we briefly review the non-relativistic
Jacobi matrix approach introduced in Refs.\cite{Hell74a,Hell74b}
and extended in \cite{matem}.
We recall only the case when the
potential vanishing faster than the Coulomb one is involved,
as we shall formulate the relativistic
formalism for this kind of potential.
The Coulomb case is much more complicated and it will
be considered elswere.
Let $\{ \phi_{n}^{l} \}{}_{n=0}^{\infty}$ be either
Laguerre or Gaussian (Hermite) basis set.
The explicit forms of both bases as well as
some other formulas concerning non-relativistic problem
(see Ref. \cite{matem}) are collected in table I.
Only the second basis, i.e., the Gaussian one, forms an orthogonal set,
hence, in general, the notion
of biorthonormality is needed.
The set $\{ \bar{\phi}_{n}^{l} \}{}_{n=0}^{\infty}$ is biorthonormal
to $\{ \phi_{n}^{l} \}{}_{n=0}^{\infty}$ with respect to the unitary
scalar product
if $\langle \bar{\phi}_{m}^{l}|\phi_{n}^{l}\rangle \equiv \int_{0}^{\infty}
\bar{\phi}_{m}^{l}(\lambda r) \phi_{n}^{l}(\lambda r) dr = \delta_{mn}$.
Biorthonormal basis functions $\{ \bar{\phi}_{n}^{l} \}$ are also given in
table I.
The important feature of the sets $\{ \phi_{n}^{l} \}$
is that the radial kinetic energy operator:
\begin{equation}
H_{0} - \frac{k^{2}}{2}\equiv-\frac{1}{2}{{\rm d}^2 \over {\rm d}r^2}+{l(l+1) \over 2 r^2}
-\frac{k^{2}}{2}
\end{equation}
if expanded in any of them, takes the tridiagonal or Jacobi form:
\begin{equation}
J_{mn}\equiv \langle\phi_{m}^{l}|( H_{0} - k^{2}/2)\phi_{n}^{l}\rangle, \\
\ \ J_{mn}\neq 0 \ \  \mbox{only for} \ \ m=n, n\pm 1.
\label{rekur}
\end{equation}
In the above $k$ is a wave number related to the energy ${\cal E}$ and mass
$m$ of the projectile
\begin{equation}
k^{2}=\frac{2m{\cal E}}{\hbar^{2}} .
\label{energian}
\end{equation}
It must be stressed here that the matrix elements
$J_{mn}$ are functions of $k$,
i.e. $J_{mn}=J_{mn}(k)$.
The regular solution $S(k,r)$ of the equation
\begin{equation}
(H_{0} - k^{2}/2)S(k,r)=0
\label{sinus}
\end{equation}
is simply proportional to the Riccati-Bessel function, satisfying
$S(k,r)\sim r^{l+1}$ as $r \rightarrow 0$ and
$S(k,r)\stackrel{r \rightarrow \infty}{\longrightarrow}
\sin(kr -\frac{\pi l}{2})$.
Using an expansion of $S(k,r)$ in the basis $\{ \phi^{l}_{n} \}$, i.e.
$S(k,r)=\sum_{n=0}^{\infty} s^{l}_{m}\phi_{n}^{l}(\lambda r)$,
one can write equation (\ref{sinus}) in the form
\begin{equation}
\sum_{n=0}^{\infty}J_{mn}s^{l}_{n}=0 .
\label{rekur1}
\end{equation}
As shown in Ref. \cite{matem}, using the explicit form of the matrix
elements $J_{mn}$ one can find the
expansion coefficients $s^{l}_{n}$ in terms of Gegenbauer
polynomials (see table I). Again we have $s_{n}^{l}=s_{n}^{l}(k)$.
In the J-matrix method to solve a scattering problem one introduces
the second, cosine-like function $C(k,r)$,
which is required to satisfy $C(k,r)\sim r^{l+1}$ as $r \rightarrow 0$ and
$C(k,r)\stackrel{r \rightarrow \infty}{\longrightarrow}
\cos(kr -\frac{\pi l}{2})$. It cannot be the second solution
of the original, homogeneous problem as this solution,
proportional to the Riccati-Neumann function, is singular at the origin.

The required $C(k,r)$ function has been found \cite{Hell74a,Hell74b,matem},
in another way, namely by solving an inhomogeneous equation :
\begin{equation}
(H_{0} - k^{2}/2)C(k,r)=\beta\bar{\phi}^{l}_{0}(\lambda r) ,
\ \ \beta=-\frac{k}{2 s^{l}_{0}}
\label{cosinus}
\end{equation}
with $s^{l}_{0}$ being the first expansion coefficient of sine solution.
Then the expansion coefficients of $C(k,r)$ satisfy the equation
\begin{equation}
\sum_{n=0}^{\infty}J_{mn}c^{l}_{n}=\beta\bar{\phi}^{l}_{0}.
\label{rekur2}
\end{equation}
The corresponding coefficients $c^{l}_{n}=c^{l}_{n}(k)$ (see table I)
have been also found \cite{matem} by some differential technique .
The calculated expansions
$S(k,r)=\sum_{n=0}^{\infty}s^{l}_{n}\phi^{l}_{n}(\lambda r)$
and $C(k,r)=\sum_{n=0}^{\infty}c^{l}_{n} \phi^{l}_{n}(\lambda r)$
have been used in an approximate solution
of the original scattering problem on
the radial potential $V=V(r)$ vanishing faster
then the Coulomb potential:
\begin{equation}
(H_{0} + V - \frac{k^{2}}{2})\psi_{E}=0 .
\label{problem}
\end{equation}
Namely, the potential $V$ has been replaced by a truncated potential
operator
\begin{equation}
V^{N}=P_{N}^{\dagger}V P_{N} ,
\end{equation}
where $P_{N}$ is the generalised projection operation:
\begin{equation}
P_{N}=\sum_{n=0}^{N-1} |\phi^{l}_{n}\rangle\langle\bar{\phi}^{l}_{n}|.
\end{equation}
The new potential operator can be written in the basis $\{ \phi^{l}_{n} \}$
as an $N \times N$ matrix with the matrix elements
$V^{N}_{mn}=\langle\phi^{l}_{n}|V\phi^{l}_{n}\rangle $.
Then the {\it exact} solution $\psi^{N}_{E}$ of the new problem:
\begin{equation}
(H_{0} + V^{N} - \frac{k^{2}}{2})\psi^{N}_{E}=0
\label{nonrelat}
\end{equation}
has been expanded in the basis $\{ \phi_{n}^{l} \}$ as
\begin{equation}
\psi^{N}_{E}(r)=\sum_{n=0}^{N-1}a^{l}_{n}\phi_{n}^{l} +
\sum_{n=N}^{\infty}(s^{l}_{n} + \tan\delta_{N}c^{l}_{n})\phi_{n}^{l}
\label{approx}
\end{equation}
to satisfy the boundary requirement
$\psi^{N}_{E}(r)\stackrel{r \rightarrow \infty}{\longrightarrow}
\sin(kr -\frac{\pi l}{2})+ \tan\delta_{N} \cos(kr -\frac{\pi l}{2})$.
The  $\tan\delta_{N}$ is an approximation of the tangent of the
sought phase shift $\delta $ of the exact solution
$\psi_{E}$ of the problem (\ref{problem}).
The left-hand side projection of (\ref{nonrelat}) onto the basis
$\{ \phi^{l}_{n}\}$ gives then infinitely many equations depending on $n$.
However all equations for $n\geq N+1$ are satisfied automatically
as coefficients $s^{l}_{n}$, $c^{l}_{n}$ satisfy the same recursion relation
(\ref{rekur1}) for any $m>0$.
The remaining finite set on equations involve $N+1$
unknowns $\tan\delta_{N}$, $\{ a_{nl} \}_{n=0}^{N-1}$.
Those equations can be easily solved \cite{Hell74a,Hell74b}.
In particular, using the recursion relation for matrix elements $J_{nm}$
the tangent can be calculated giving
\begin{equation}
\tan\delta_{N}=-{s^{l}_{N-1} + g_{N-1,N-1}({\cal E}) J_{N,N-1} s^{l}_{N} \over
c^{l}_{N-1} + g_{N-1,N-1}({\cal E}) J_{N,N-1} c^{l}_{N}}
\label{nrshift}
\end{equation}
where $g_{N-1,N-1}({\cal E})=\sum_{n=0}^{N-1}\Gamma^{2}_{N-1,m}/({\cal E}_m -{\cal E})$ with
the matrix $\Gamma$ diagonalising the finite-dimensional problem
$( {\Gamma}^{\dagger}P^{\dagger}(H_{0} + V - \frac{k^{2}}{2}) P\Gamma )_{mn}=
({\cal E}_n-{\cal E})\delta_{mn} $.
Here the energy dependent quantity
$g_{N-1,N-1}({\cal E})$ can be viewed as the
matrix element of the inverse of
the truncated operator $P^{\dagger}(H_{0} + V^{N} - \frac{k^{2}}{2})P$
if restricted to the $N$-dimensional space where it
does not vanish.
The quantities ${\cal E}_n$
\cite{Harris} (see also \cite{Yamani} and references
therein).
\section{Relativistic Jacobi-matrix problem}

Now we shall turn to the relativistic problem.
Before the formulation of the method we shall find the relativistic
counterparts of $S(k,r)$ and $C(k,r)$ in some suitable basis.
We shall also calculate the relativistic Jacobi matrix elements
in this basis.
For this purpose consider the free Dirac equation:
\begin{eqnarray}
({\cal H}_{0} - E/c \hbar)\Psi\equiv
  \left( \begin{array}{cc}
         (mc^2 - E)/c \hbar & - d/dr + \kappa /r  \\
          d/dr + \kappa / r &(-mc^2 - E)/c \hbar
         \end{array}
  \right)
  \left( \begin{array}{c}
           F(r) \\
           G(r)
         \end{array}
  \right)=
  \left( \begin{array}{c}
           0 \\
           0
         \end{array}
  \right)
\label{Diracf}
\end{eqnarray}
In the above the total energy $E$ is related
to the rest energy ${\cal E}$ as $E={\cal E}+mc^{2}$.
Let $l(\kappa)$ be the non-negative solution
of the equation $l(l+1)=\kappa(\kappa+1)$, i.e.
$l(\kappa)=\kappa $ and $l(\kappa)=-\kappa-1$
for positive and negative $\kappa$, respectively.
We shall usually omit the symbol $\kappa$ in the notation throughout the text
and write  $l$ only, remembering that the latter depends on $\kappa$.
Then equation (\ref{Diracf})
has two independent solutions.
The first one, regular at the origin
\begin{eqnarray}
 \Psi_{reg}(r) = \left( \begin{array}{c}
     F_{reg}(r)    \\
     G_{reg}(r)
   \end{array} \right ) \sim
   \left( \begin{array}{c}
    \hat{\jmath}_{l}(\tilde{k}r)     \\
   \pm \epsilon \hat{\jmath}_{l \mp 1}(\tilde{k}r)
   \end{array} \right ) .
\label{regrel}
\end{eqnarray}
is constituted by the Riccati-Bessel functions with boundary behaviour
\begin{equation}
        \hat{\jmath}_{l}(x) \stackrel{x \rightarrow 0}{\longrightarrow} \frac{x^{l+1}}{(2l+1)!!}
       \ \   \mbox{and}   \ \
        \hat{\jmath}_{l}(x) \stackrel{x \rightarrow \infty}{\longrightarrow} \sin(x-{\pi l \over 2})
\label{Bessel}
\end{equation}
The numbers $\epsilon $ and $\tilde{k}$ in (\ref{regrel})
are standard abbreviations
\begin{eqnarray}
\epsilon \equiv \sqrt{\frac{E-mc^2}{E+mc^2}}, \ \ \
\tilde{k}\equiv {\sqrt{(E-mc^2)(E+mc^2)} \over c \hbar}.
\label{5}
\end{eqnarray}
The quantity $\tilde{k}$ converges in the non-relativistic
limit $c \rightarrow \infty$ to the number $k=\sqrt{\frac{2m{\cal E}}{\hbar^{2}}}$.
The second solution of (\ref{Diracf}), irregular at zero is given by
\begin{eqnarray}
\Psi_{irr}(r) = \left( \begin{array}{c}
     F_{irr}(r)    \\
     G_{irr}(r)
   \end{array} \right )\sim
   \left( \begin{array}{c}
    -\hat{n}_{l}(\tilde{k}r)     \\
   \pm \epsilon \hat{n}_{l\mp 1}(\tilde{k}r)
   \end{array} \right ) .
\label{irregrel}
\end{eqnarray}
Here we have the Ricatti-Neumann functions with properties:
\begin{equation}
        \hat{n}_{l}(x) \stackrel{x \rightarrow 0}{\longrightarrow}
        -\frac{(2l-1)!!}{x^l}
       \ \   \mbox{and}   \ \
        \hat{n}_{l}(x) \stackrel{x \rightarrow \infty}{\longrightarrow}
       - \cos(x-{\pi l\over 2})
\label{Neumann}
\end{equation}
In both solutions (\ref{regrel}) and (\ref{irregrel})
the upper and lower signs in the
small components correspond to negative and positive $\kappa$, respectively.
From the above it can be
immediately seen that the regular solution $\Psi_{reg}$
is the relativistic counterpart of
non-relativistic function
$S(k,r)$.
For the sake of consistency with the non-relativistic
case, hereafter
we shall denote the relativistic sine-like solution
$\Psi_{reg}$ by $\Psi_{S}$. To develop the Jacobi matrix
analysis we have to introduce a suitable basis set.
\subsection{The basis set}
in the Hilbert space
$L^{2}(0,\infty) \otimes C^{2}$ on which the
Dirac operator from (\ref{Diracf}) is defined.
Let again $\{ \phi^{l}_{n}(x) \}$ be either the Laguerre or the Gaussian
basis set  and let $ \psi^{l}_{n}(\lambda r)=
(\kappa/r+{\rm d}/{\rm d}r)\phi^{l}_{n}(\lambda r)$\footnote{
One should keep in mind that here dependence on $\kappa$
is not only present via $l$ coefficient, but
via operator $(\kappa/{\rm r}+{\rm d}/{\rm dr})$.}
Then the basis set defined for our purposes is
\begin{equation}
\Phi_{n \kappa}^{+}(r) \equiv
\left(\begin{array}{c} \phi^{l}_{n}(\lambda r) \\ 0
\end{array}\right), \ \ \
\Phi_{n \kappa}^{-}(r) \equiv
\left(\begin{array}{c} 0 \\ \psi^{l}_{n}(\lambda r)
 \end{array}\right)
\label{Baza}
\end{equation}
The above set depends on the positive reals number $\lambda$
which can be treated as a nonlinear variational
parameter (see, for instance, \cite{GoldPRA}.
Note that the set (\ref{Baza}) satisfies the ``kinetic balance condition''.
The latter condition is generally defined as a requirement
that, if the funcitions $\{ \gamma_{i} \}$ are used
to expand large component of solution of Dirac equation,
then the basis $\{ \omega_{i} \}$ used for expansion of
small component should consist of linear combinations
of functions $\{ (\kappa /r + {\rm d}/{\rm d}r)\gamma_{i} \}$.
Use of such a basis is the simplest way \cite{Dyall1,Dyall2}
to omit the problem of so called ``finite basis set disease''
(see \cite{Schw1,Schw2,Kutz})
in estimation of bound states of the atomic system.
It seems that it would be also interesting in future to
consider the relativistic J-matrix problem in
the context of the relativistic
Sturmian basis (see \cite{Radek} and references therein)
as it is known that the relativistic free particle Green function
takes particularily simple form in this basis.

The biorthonormal elements to the functions (\ref{Baza}) obviously are
$\bar{\Phi}_{n \kappa}^{+}(r) = (\bar{\phi}^{l}_{n}(\lambda r)
, 0 ) ^{T}$, $\bar{\Phi}_{n \kappa}^{-}(r) = (0,
\bar{\psi}^{l}_{n}(\lambda r))^{T}$.
As usual, we denote by $\bar{f_{n}}$ the element biorthonormal to $f_{n}$.
The elements $\bar{\phi}^{l(\kappa)}_{n}$  \cite{matem}
are recalled in table I.
Here we shall calculate the biorthonormal elements
$\bar{\psi}^{l(\kappa)}_{n}$.

It is easy to show integrating by parts, that
biorthonormal elements
$\{\bar{\psi}^{l}_n(x)\}^{\infty}_{n=0}$ should satisfy
the equation
\begin{equation}
\int_{0}^{\infty}\left[\left(\frac{\kappa}{r}-\frac{{\rm d}}{{\rm d}r}\right)
\bar{\psi}^{l}_{m}(\lambda r)\right]
\phi^{l}_{n}(\lambda r)
d( r)=\delta_{mn} .
\label{7a}
\end{equation}
Hence it suffices only to solve the following inhomogeneous differential
equation
\begin{equation}
\left({\kappa\over x}-{{\rm d}\over {\rm d}x}\right)\bar{\psi}^{l}_{n}(x)=
\bar{\phi}^{l}_{n}(x), \ \ x=\lambda r.
\label{9a}
\end{equation}

The resulting functions are given in Table II.
They all belong to the space $L^{2}(0,\infty)$. This fact is obvious
apart from the case of negative $\kappa$ for the Gaussian set.
This case needs more careful analysis as here it
is not possible to give the functions by explicit
formula. For all $\kappa<0$ the functions $\{ \bar{\psi}_{n}^{l} \}$ due
to Gaussian set behave as $r^{l+2}$ at the origin, and vanish at infinity
not slower than  $r^{-(l+1)}$ as the limit of the occurring integral is finite.
Then $(\psi_{n}^{l})^{2}$ behaves for $r \rightarrow \infty $ as $r^{-2(l+1)}$
and, as $l$ is nonnegative, $||\psi_{n}^{l}||$ exists.
Thus in both cases, when $\{ \phi^{l}_{n} \}$ is
either the Laguerre or the Gaussian basis set,
all elements $\{ \bar{\psi}^{l}_{n}  \}$
biorthonormal to new functions
$\psi^{l}_{n}=(\kappa/r+d/dr)\phi^{l}_{n} $ belong to
$ L^{2}(0,\infty) $. Then obviously biorthonormal elements
$\bar{\Phi}_{n\kappa}$ due to the relativistic
case belong to the Hilbert space
$L^{2}(0,\infty) \otimes C^{2}$.
Note that we do {\it not} need the explicit forms of
biorthonormal functions in our considerations.

\subsection{Expansions of relativistic sine and cosine solutions}
Now we are in the position to find the expansions
of sine-like $\Psi_{S}=\Psi_{reg}$ and cosine-like $\Psi_{C}$ solutions.
For the latter we demand to satisfy three requirements:

(1) $\Psi_{C}$ should have $\Psi_{irr}$ type asymptotic
form,

(2)  $\Psi_{C}$ should exhibit regular behaviour at the origin,

(3) coefficients of $\Psi_{C}$ expansion should satisfy
(apart from at most the few first ones) the same recurrence
equations like the ones of $\Psi_{reg}=\Psi_{S}$.

Consider first the solution $\Psi_{U}(\tilde{k},r)= \left ( \begin{array}{cc}
F_{U}(\tilde{k},r), & G_{U}(\tilde{k},r) \end{array} \right )^{T} $
of the inhomogeneous equation of type (\ref{Diracf}):
\begin{equation}
({\cal H}_{0} - \frac{E}{c \hbar})\Psi_{U}\equiv \Phi_{inh} .
\label{inho}
\end{equation}
In the above the index $U=S, C$ corresponds to sine-line and cosine-like
solution.
The inhomogeneity is chosen as
$\Phi_{inh}=\Phi^{+}_{0\kappa}=
\left ( \begin{array}{cc}\Omega_{U}\bar{\phi}_{0}^{l}, & 0 \end{array} \right )^{T}$
and the coefficients $\Omega_{U}, U=S,C $ are $\Omega_{S}=0$,
$\Omega_{C}=-\epsilon /s^{l}_{0}$.

Equation (\ref{inho}) can be also written as
\begin{eqnarray}
   \begin{array}{c}
       (\kappa / r - {\rm d}/{\rm d} r) G_{U}-\tilde{k}\epsilon F_{U} = \Omega_{U} \bar{\psi}_{0}^{l} \\
       (\kappa / r + {\rm d}/{\rm d} r)F_{U}-\frac{\tilde{k}}{\epsilon}G_{U} = 0 .
   \end{array}
\label{Diracf1}
\end{eqnarray}
We can introduce the relativistic counterpart of the
Jacobi matrix :
\begin{equation}
{\cal J}_{mn}^{ss'}\equiv
\langle \Phi_{m \kappa}^{s}|( {\cal H}_{0} - E/ c\hbar)
\Phi_{n \kappa}^{s} \rangle,\ s, s'=\pm, \ m, n=0, 1, 2, ...
\label{Jm}
\end{equation}

The matrix elements of ${\cal J}$
can be expressed in an extremely simple form.
To see this, let us define the $2 \times 2 $ matrices
${\cal J}_{mn}$ defined by their matrix elements as
$\{ {\cal J}_{mn} \}_{ss'} \equiv {\cal J}_{mn}^{ss'}$.
Then it can be easily seen that
in the spinor basis the new matrix takes
the particularly simple form:
\begin{equation}
{\cal J}_{mn}=
\left (
\begin{array}{cc}
-\tilde{k}\epsilon \langle\phi_{m}^{l}|\phi_{n}^{l} \rangle &
                \langle\psi_{m}^{l}|\psi_{n}^{l} \rangle \\
                \langle\psi_{m}^{l}|\psi_{n}^{l} \rangle &
              -{\tilde{k} \over \epsilon} \langle \psi_{m}^{l}|\psi_{n}^{l} \rangle
\end{array}
\right ) .
\label{Jm1}
\end{equation}
The explicit forms of the integrals
constituting elements of the above matrix are given in table III.
They are simply related to the
non-relativistic J-matrix elements (\ref{rekur})(c.f. \cite{matem}):
\begin{equation}
J_{mn}={1 \over 2} \langle \psi_{m}^{l} | \psi_{n}^{l} \rangle
-{k^{2} \over 2} \langle \phi_{m}^{l} | \phi_{n}^{l} \rangle .
\label{wzorek}
\end{equation}
Now we shall predict the expansions of
the two solutions in basis (\ref{Baza}) in the following form
\begin{equation}
\Psi_{U}= \sum_{s=\pm} \sum_{n=0}
^{\infty} u_{n\kappa}^{s}\Phi_{n\kappa}^{s}
\equiv \sum_{n=0}^{\infty} u_{n}^{l}(\tilde {k})
 \left( \begin{array}{c}
     \phi_{n}^{l}    \\
     (\epsilon /\tilde{k})\psi_{n}^{l}
   \end{array} \right ) , \ \ U=S,C; \ \  u=s,c \\
\label{exp}
\end{equation}
i.e. we predict that large components of sine-like and cosine-like
solutions are given by the same expansion coefficients
$s_{n}^{l}, c_{n}^{l}$ as in the non-relativistic case, only taken
in the modified point $\tilde{k}$
and that the small
components coefficients
are only rescaled by $\epsilon /\tilde{k}$.

It can be easily verified that (\ref{exp})
really solves the equation (\ref{Diracf}).
Namely putting the above expansion into the equation
and using the definition of matrix elements (\ref{Jm})
we get the infinite set of equations:
\begin{equation}
\sum_{s'=\pm} \sum_{n=0}^{\infty}{\cal J}^{ss'}_{mn}
u_{n\kappa}^{s'}=\Omega_{U}\bar{\phi}_{0}^{l}\delta_{m0}\delta_{s,+},
\ \ s=\pm, m= 0, 1, 2, ...
\label{equ}
\end{equation}
as for any pair ${m, n}$ fixed the second element in lower
row of the matrix (\ref{Jm1}) is rescaled by $-\frac{\tilde{k}}{\epsilon}$
we obtain immediately that all equations (\ref{equ})
with a negative ``index'' $s=$``-'' are satisfied trivially.
Recalling the definition
of $l=l(\kappa)$ (cf. the remark following
equation (\ref{Diracf})) after integration by parts one gets
$\langle\psi_{m}^{l}|\psi_{n}^{l}\rangle=
\langle(\frac{\kappa}{r} + \frac{d}{dr})\phi_{m}^{l}|(\frac{\kappa}{r} + \frac{d}{dr})\phi_{n}^{l}\rangle=
\langle\phi_{m}^{l}|(-\frac{d^{2}}{dr^{2}}+\frac{l(l+1)}{r^{2}})\phi_{n}^{l}\rangle$.
Taking into account the form of the upper row of the
matrix (\ref{Jm1}) and the identity
(\ref{wzorek}) we obtain immediately that
equations (\ref{equ}) with the ``index'' $s=$``+'' have the identical
form with the sets of equations (\ref{rekur1}), (\ref{rekur2}) if
the latter are evaluated at $\tilde{k}$ instead of $k$.
Thus we have shown that the expansions
(\ref{exp}) are in fact solutions of equation (\ref{Diracf1}).
From the non-relativistic case we see that their large components
have the desired behaviour at the origin and at infinity.
Moreover, as the equations of the type (\ref{Diracf})
are coupled and the behaviour of the one component determines
the behaviour of the other one.
Hence both the components of
the solutions $\Psi_{S}$,  $\Psi_{C}$ have the asymptotic
behaviour we need for purposes of our method, i.e.,
$\Psi_{S}$ is simply the regular solution,
$\Psi_{C}$ behaves as $\Psi_{irr}$ at infinity and
as $\Psi_{reg}$  at the origin. As the inhomogeneity
involves only one biorthonormal element $\Psi_{0\kappa}^{+}$ both functions
satisfy {\it the same} set of equations apart from the first one
(see formula (\ref{Jm1})).
\section{Potential scattering}

Now we shall consider the central problem of our
paper which is the approximate solution within
the ${\cal J}$-matrix formalism. Consider again the radial part of
the scattering problem of a projectile on a target described
by a sufficiently regular potential $V= V(r)$ vanishing
at infinity faster then the Coulomb potential.
To solve the problem one has to find the solution of the following equation
\begin{equation}
\left( {\cal H}_{0} + \frac{V}{c\hbar} - \frac{E}{c\hbar} \right) \Psi_{E}=0.
\end{equation}
A solution $\Psi_{E}$ of the above equation
is required to satisfy the boundary condition
$\Psi_{E}(\tilde{k},r)
\mathop{\sim}\limits^{r \rightarrow \infty}
\Psi_{S}(\tilde{k},r) + \tilde{t}\Psi_{C}(\tilde{k},r)$
where the tangent of the phase shift, $\tilde{t}=\tan\tilde{\delta}$,
is to be found.

To develop the formalism of the relativistic J-matrix (we shall
denote it by ${\cal J}$-matrix to distinguish from the non-relativistic case)
we use the generalised projection operators:
\begin{equation}
{\cal P}_{N}=
\sum_{s=\pm}\sum_{n=0}^{N-1}
|\Phi_{n \kappa}^{s}\rangle \langle \bar{\Phi}_{n \kappa}^{s}| ,
\end{equation}
and introduce the truncated potential:
\begin{equation}
{\cal V}^{N}={\cal P}_{N}^{\dagger} \frac{V}{c\hbar}{\cal P}_{N} .
\end{equation}
where ${\cal P}_{N}^{\dagger}$ corresponds to
hermitian conjugate of ${\cal P}_{N}$.

Now one can seek the exact solution of the equation with truncated potential:
\begin{equation}
\left( {\cal H}_{0} + {\cal V}^{N} - \frac{E}{c\hbar} \right) \Psi^{N}_{E}(r)=0.
\label{approxrel}
\end{equation}
Note that for any $\psi \in L^{2}(0,\infty) \otimes C^{2} $
the function ${\cal V}^{N}\psi$ vanishes at infinity faster then ${1 \over r}$.
Recall that we assumed that our original potentials vanish
at infinity faster then $\frac{1}{r^{2}}$.
Thus although equation (\ref{approxrel})
has not a standard Dirac equation form with
the same scalar potential in its large and small part,
still its solution asymptotically satisfies
{\it free} Dirac equation.
Hence the solution  $\Psi_{E}^{N}$
satisfies the boundary condition
\begin{equation}
\Psi^{N}_{E}(\tilde{k},r) \sim \Psi_{S}(\tilde{k},r) + \tilde{t}_{N}\Psi_{C}(\tilde{k},r) ,
\label{asymptapp}
\end{equation}
where $\tilde{t}_{N}$ is an approximated tangent of phase shift.
As the potential operator
${\cal V}^{N} \stackrel{N \rightarrow \infty}{\longrightarrow}
\frac{V}{c\hbar}$
we expect that for $N \rightarrow \infty$,  $\tilde{t}_{N}$ converges
to correct value $\tilde{t}=\tan\tilde{\delta}$.

Now we shall find more details about the form of the solution
$\Psi^{N}_{E}$. The most general formula is
\begin{equation}
\Psi^{N}_{E}=
\sum_{s=\pm} \sum_{m=0}^{\infty} d_{m\kappa}^{s} |\Phi_{m \kappa}^{s} \rangle .
\end{equation}
Consider the matrix representation of
equation (\ref{approxrel}). Putting the expansion of
the function $\Psi_{E}^{N}$ in the basis $\{ \Psi_{n\kappa}^{\pm} \}$
we get the infinite set of equations:
\begin{equation}
\sum_{s'=\pm} \sum_{n=0}^{\infty} ( {\cal J} + {\cal V}^{N})^{ss'}_{mn}
d_{n\kappa}^{s'}=0, \ \ s=\pm, m= 0, 1, 2, ...
\label{equd}
\end{equation}
It can be easily seen
by the right-hand side projection of equations
(\ref{Diracf1}) onto the basis
$\{ \Psi_{n\kappa}^{\pm} \}$.

According to analysis following
the formula (\ref{equ}) the expansion coefficients $\{ d_{n\kappa}^{\pm} \}$
of $\Psi^{N}_{E}$ must satisfy:
(a) for the large component $\sum_{n=0}^{\infty}J_{mn}(\tilde{k})d_{n\kappa}^{+}=0$,
$m>N$ with elements $J_{mn}(\cdot)$ given by the non-relativistic formula,
(b) for the small component $d_{n\kappa}^{-}=\frac{\epsilon}{\tilde{k}}d_{n\kappa}^{+}$.
Moreover we impose the additional condition (c) $F^{N}_{E} \sim
S(\tilde{k},r)+ \tilde{t}_{N} C(\tilde{k},r)$ (see condition (\ref{asymptapp})).
This gives us, together with the condition (b),
the following required form of the sought
solution $\Psi^{N}_{E}$
of equation (\ref{approxrel}) (c.f. \cite{matem}):
\begin{equation}
\Psi^{N}_{E} =
\sum_{m=0}^{N-1}
\left( \begin{array}{c}
d_{m\kappa}^{+} \phi_{m}^{l} \\
d_{m\kappa}^{-} \frac{\epsilon}{\tilde{k}}\psi_{m}^{l}
\end{array} \right )+
\sum_{m=N}^{\infty}
\left( \begin{array}{c}
(s_{\kappa m}^{+} + \tilde{t}_{N}c_{\kappa m}^{+}) \phi_{n}^{l} \\
(s_{\kappa m}^{-} + \tilde{t}_{N}c_{\kappa m}^{-})
\psi_{n}^{l}
\end{array} \right ),
\label{approxrozw}
\end{equation}
where the abbreviations $s^{\pm}_{\kappa m}$,
$c^{\pm}_{\kappa m}$ has been used according to (\ref{exp}).
After adding and subtracting the term
$\sum_{m=0}^{N-1}
\left (\begin{array}{cc}
(s_{\kappa m}^{+} + \tilde{t}_{N}c_{\kappa m}^{+})
\phi_{n}^{l}, &
(s_{\kappa m}^{-} + \tilde{t}_{N}c_{\kappa m}^{-})\psi_{n}^{l}\end{array} \right)^{T}$
to the left hand side of the above
equation it is straightforward to see that the above function
satisfies the asymptotic condition (\ref{asymptapp}).

Let us turn back to equations (\ref{equd}).
In general, in analogy to the non-relativistic case,
they can be schematically represented as follows:
{\scriptsize
\begin{displaymath}
\left( \begin{array}{cccccccccccccccccc}

  \cdot&\cdot&\cdot&&&&&&&&&&&&&\cdot&\cdot&\cdot\\
  &X&X&X&&&&&0&& &&&&X&X&X& \\
  &&X&X&X&&&&&& &&&X&X&X&& \\
  &&&X&X&X&X&X&X&& &&X&X&X&&& \\
  &&&&X&X&X&X&X&& &X&X&X&&&& \\
  &&&&X&X&X&X&X&&X&X&X&&&&& \\
  &&&&X&X&X&X&X&X&X&X&&&&&& \\
  &&&&X&X&X&X&X&X&X&&&&&&0& \\
  &0&&&&&&X&X&X&X&X&X&X&&&& \\
  &&&&&&X&X&X&X&X&X&X&X&&&& \\
  &&&&&X&X&X&&X&X&X&X&X&&&& \\
  &&&&X&X&X&&&X&X&X&X&X&&&& \\
  &&&X&X&X&&&&X&X&X&X&X&X&&& \\
  &&X&X&X&&&&&& &&&X&X&X&& \\
  &X&X&X&&&&&&0& &&&&X&X&X& \\
 \cdot &\cdot&\cdot&&&&&&&&&&&&&\cdot&\cdot&\cdot
   \end{array}
\right)
\left( \begin{array}{c}
     \cdot \\
       s_{N+1,l}^{+} \\
       s_{N,l}^{+} \\
       d_{N-1,l}^{+}  \\
       \cdot \\
        \cdot\\
       d_{1,l}^{+} \\
       d_{0,l}^{+} \\
       d_{0,l}^{-} \\
       d_{1,l}^{-} \\
       \cdot \\
       \cdot \\
       d_{N-1,l}^{-} \\
       s_{N,l}^{-} \\
       s_{N+1,l}^{-} \\
    \cdot \\
   \end{array}
\right)=
\left( \begin{array}{c}
       \cdot \\
       0 \\
       0 \\
       0 \\
       \cdot \\
       \cdot \\
       0 \\
       0 \\
       0 \\
       0 \\
       \cdot \\
       \cdot \\
       0 \\
       0 \\
       0 \\
       \cdot \\
   \end{array}
\right)
\end{displaymath}}

From the construction of the required form
(\ref{approxrozw}) we see that
all the equations for $m>N$ are satisfied automatically.
Thus one has to solve the remaining $2N+2$ equations
with the unknowns $\tilde{t}_{N}$,
$d_{0\kappa}^{+},d_{1\kappa}^{+},..., d_{N-1,\kappa}^{+};
d_{0\kappa}^{-},d_{1\kappa}^{-},..., d_{N-1,\kappa}^{-}$.
Note that here the number of equations is greater
then the number of sought quantities ($2N+1$), so in general
the set of equations of such a form can have no solution.
But in our particular case
the solution certainly exists
as the general theory of differential equations
assures the existence of $\Psi_{E}^{N}$ and, according
to the previous analysis,
(\ref{approxrozw}) represents the most general required
form of $\Psi_{E}^{N}$.

Using equations (\ref{equd})
one obtains the following form of the remaining equations:

{\scriptsize
\begin{displaymath}
 \left( \begin{array}{ccccccc}
-{\cal J}^{++}_{N,N-1}c_{N-1}^{+}& {\cal J}^{++}_{N,N-1} & 0 &\cdot \cdot \cdot &
0 & {\cal J}^{+-}_{N,N-1} & -{\cal J}^{+-}_{N,N-1}c_{N-1}^{-} \\
{\cal J}^{++}_{N-1,N}c_{N}^{+}&({\cal J}+{\cal V}^{N})_{N-1,N-1}^{++}&
({\cal J}+{\cal V}^{N})_{N-1,N-2}^{++} &\cdots&
{\cal J}_{N-1,N-2}^{+-}&
{\cal J}_{N-1,N-1}^{+-}&{\cal J}^{+-}_{N-1,N}c_{N}^{-} \\
0&({\cal J}+{\cal V}^{N})_{N-2,N-1}^{++}&
({\cal J}+{\cal V}^{N})_{N-2,N-2}^{++} &\cdots&
{\cal J}_{N-2,N-2}^{+-}&
{\cal J}_{N-2,N-1}^{+-}&0 \\
\vdots&\vdots&\vdots&\cdots&\vdots&\vdots&\vdots \\
0&({\cal J}+{\cal V}^{N})_{0,N-1}^{++}&
({\cal J}+{\cal V}^{N})_{0,N-2}^{++} &\cdot \cdot \cdot&
({\cal J}+{\cal V}^{N})_{0,N-2}^{+-}&
({\cal J}+{\cal V}^{N})_{0,N-1}^{+-}&0 \\
0&({\cal J}+{\cal V}^{N})_{0,N-1}^{-+}&
({\cal J}+{\cal V}^{N})_{0,N-2}^{-+} &\cdot \cdot \cdot&
({\cal J}+{\cal V}^{N})_{0,N-2}^{--}&
({\cal J}+{\cal V}^{N})_{0,N-1}^{+-}&0 \\
\vdots&\vdots&\vdots&\cdots&\vdots&\vdots&\vdots \\
0&{\cal J}_{N-2,N-1}^{-+}&
{\cal J}_{N-2,N-2}^{-+} &\cdot \cdot \cdot&
({\cal J}+{\cal V}^{N})_{N-2,N-2}^{--}&
({\cal J}+{\cal V}^{N})_{N-2,N-1}^{--}&0 \\
{\cal J}^{-+}_{N-1,N}c_{N}^{+}
&{\cal J}_{N-1,N-1}^{-+}&
{\cal J}_{N-1,N-2}^{-+} &\cdot \cdot \cdot&
({\cal J}+{\cal V}^{N})_{N-1,N-2}^{--}&
({\cal J}+{\cal V}^{N})_{N-1,N-1}^{--}&{\cal J}^{--}_{N-1,N}c_{N}^{-} \\
-{\cal J}^{-+}_{N,N-1}c_{N-1}^{+}&{\cal J}^{-+}_{N,N-1}&0&\cdot \cdot \cdot&
0&{\cal J}^{--}_{N,N-1}
&-{\cal J}^{-}_{N,N-1}c_{N-1}^{-}
\end{array}
\right)
\end{displaymath}}
{\scriptsize
\begin{displaymath}
\times \left( \begin{array}{c}
         \tilde{t}_{N} \\
         d_{N-1}^{+} \\
         d_{N-2}^{+} \\
 \cdot \\
 d_{0}^{+} \\
 d_{0}^{-} \\
 \cdot \\
 d_{N-2}^{-} \\
 d_{N-1}^{-} \\
         \tilde{t}_{N}
\end{array} \right ) =
\left( \begin{array}{c}
{\cal J}_{N,N-1}^{++}s_{N-1}^{+}
+{\cal J}_{N,N-1}^{+-}s_{N-1}^{-}\\
-{\cal J}_{N-1,N}^{++}s_{N}^{+}
 -{\cal J}_{N-1,N}^{+-}s_{N}^{-}\\
0 \\
 \cdot \\
0\\
0\\
 \cdot \\
0 \\
-{\cal J}_{N-1,N}^{-+}s_{N}^{+}
-{\cal J}_{N-1,N}^{--}s_{N}^{-} \\
{\cal J}_{N,N-1}^{-+}s_{N-1}^{+}
+{\cal J}_{N,N-1}^{--}s_{N-1}^{-}
  \end{array} \right )
\end{displaymath}
}

Keeping in mind that the inner $2N \times 2N$ matrix
$({\cal J} + {\cal V}^{N})^{ss'}_{nm}$, $s=\pm$, $m,n=0,1, ... ,N-1$
is Hermitian and real, hence symmetric, and recalling
the definition of ${\cal V}^{N}$ we can solve the above equations
by some orthogonal matrix $\Gamma$ (cf. the non-relativistic case):
\begin{equation}
( \Gamma^{\dagger} {\cal P}_{n}^{\dagger}({\cal H}_{0} + \frac{V}{c\hbar}- \frac{E}{c\hbar})
{\cal P}_{N}\Gamma)_{mn}^{ss'}=\frac{1}{c\hbar}(E_{n}^{s} - E)\delta_{nm}\delta_{ss'}
\end{equation}
$2N \times 2N$ matrix ${\cal G}(E)$ with elements defined as
\begin{equation}
{\cal G}_{mn}^{ss'}(E)=\sum_{p=\pm}\sum_{i=0}^{N-1}
c\hbar\frac{\Gamma^{sp}_{m,i}\Gamma^{s'p}_{n,i}}{E_{i}^{p}-E}
\end{equation}
is an inverse of the $2N \times 2N $ matrix representation
of the truncated operator ${\cal P}_{N}^{\dagger}(
{\cal H}_{0} +\frac{V}{c\hbar} - \frac{E}{c\hbar}){\cal P}_{N} $.
It can be viewed as the approximation of the
relativistic Green function in the basis (\ref{Baza}).
The numbers $E_{n}^{p}$ are the relativistic counterparts
of the Harris eigenvalues \cite{Harris}.
They represent a finite approximation of
the spectrum of the relativistic Hamiltonian
${\cal H}_{0} + \frac{V}{c\hbar}$.
In particular, they include positive approximations
of the first $N$ energy levels due to the
potential $V$ and the $N$ negative pseudo-energies
due to the continuous spectrum.
As our basis (\ref{Baza}) satisfies the kinetic balance condition
\cite{Dyall1,Dyall2}
there is a hope that $E_{i}^{p}$ satisfy
the generalised form of the Hylleraas-Undheim theorem
(see, for instance \cite{Grant,Gold} and references therein).
It means, in particular,  that (i) N positive
values among  set $\{ E_{i}^{p}   \}$ approximate
the exact eigenenergies
${\cal H}_{0} + \frac{V}{c\hbar}$ from
the above and that (ii)
the remaining N eigenvalues have values
below $-mc^{2}$.

We can introduce now the $ (2N + 2) \times (2N + 2) $
block-diagonal matrix :

\begin{equation}
\tilde{\Gamma}_{(2N + 2) \times (2N + 2)}=
diag(1, \
 \Gamma_{2N \times 2N}, \ 1)
\end{equation}
and act with it on the left-hand side of the above
set of $ 2N + 2 $ equations.
Using the fact that the matrix ${\cal J}_{N,N-1}$ given by (\ref{Jm1})
is nonsingular the set of $(2N+2)\times (2N+2)$
the equations can be solved with respect to the approximate
tangent of phase shift.
Using the properties of the
coefficients of the matrix ${\cal J}$ one can
derive the tangent of the appoximated phase shift
in the form similar to the non-relativistic formula:
\begin{equation}
\tilde{t}_{N}=-\frac{s^{l}_{N-1}(\tilde{k})+(2\epsilon/\tilde{k}){\cal G}^{++}_{N-1,N-1}(E)
J_{N,N-1}(\tilde{k})
s^{l}_{N}(\tilde{k})}{c^{l}_{N-1}(\tilde{k})+(2\epsilon/\tilde{k}){\cal G}^{++}_{N-1,N-1}(E)
J_{N,N-1}(\tilde{k})c^{l}_{N}(\tilde{k})}.
\label{rshift}
\end{equation}
Note that in the above the $J_{N,N-1}$ stands for the
non-relativistic $J$-matrix element (see (\ref{wzorek})).
The fact that we have $ 2N+2 $ equations and $ 2N+1 $ unknows
results in second, very similar formula for $\tilde{t}_{N}$
with $(\tilde{k}/\epsilon){\cal G}^{-+}_{N-1,N-1}(E)$ instead of
${\cal G}^{++}_{N-1,N-1}(E)$.
From the previous analysis we know that both
equations must give {\it the same} $\tilde{t}_{N}$
which means that one has
${\cal G}^{-+}_{N-1,N-1}(E)=(\epsilon/\tilde{k}){\cal G}^{++}_{N-1,N-1}(E)$.
\section{Discussion}
Comparing equation (\ref{rshift}) with the non-relativistic formula
(\ref{nrshift}) one can see that apart from
the quantity $(2\epsilon/\tilde{k}){\cal G}^{++}_{N-1,N-1}(E)$,
all elements of the expression for tangent of the phase shift have the same form as in
(\ref{nrshift}), they are only evaluated in
relativistic wave number $\tilde{k}$.

Now let us note that for any $N$ the above
formula for tangent shift converges to the non-relativistic limit
as the speed of light $c$ approaches infinity.
Indeed, the used
basis (\ref{Baza}) ensures (see \cite{Dyall2,Grant})
that in the limit of infinite $c$ the large component satisfies the
correct Schr\"odinger equation (\ref{nonrelat})
with the wave number $k=\lim_{c \rightarrow \infty} \tilde{k} $.
This means that the related tangent of the phase shift must also satisfy
a correct limit, i.e.
\begin{equation}
\lim_{c \rightarrow \infty} \tilde{t}_{N} =t_{N}.
\end{equation}
From the above we get immediately
$\lim_{c \rightarrow \infty}
(2\epsilon/\tilde{k}){\cal G}^{++}_{N-1,N-1}(E)=g_{N-1,N-1}({\cal E})$.
Moreover $(2\epsilon/\tilde{k}){\cal G}^{++}_{N-1,N-1}(E)$ plays
the analogous role as $g_{N-1,N-1}({\cal E})$.
In fact, the matrices ${\cal G}(E)$ and $g({\cal E})$ can be viewed as 
the finite
approximations of the Green functions of the relativistic and non-relativistic
Hamiltonians with the potential $V$, respectively.
The form of the factor $2\epsilon/\tilde{k}$
is simply connected with the normalisations of the
Green functions in both cases.
It can be seen from the simple analysis of the set of
second order equations derived in a standard way from the Dirac equation.

From the practical point of view, the convergence can be improved
with the help of additional parameter $\lambda$.
As we mentioned before, the latter can be treated
as an additional variational parameter.
In particular its optimal value will depend on
the range of the potential.
It can be simply seen that potentials of long range
should be treated with small $\lambda$ while
potentials with support located close to the
origin will require large values of the parameter.

In conclusion, we have provided the relativistic version
of Jacobi matrix method for well defined class of 
potentials. The usage of the basis satisfying the
``kinetic balance condition'' allowed for a simple formulation of
the method. In particular, the derived expression for the
tangent of the phase shift is similar to its non-relativistic counterpart
and reproduces the latter as a correct non-relativistic limit.

The author is especially grateful to R. Szmytkowski for suggesting
the problem, many helpful discussions, comments and remarks.
He also thanks J. E. Sienkiewicz for discussion on
kinetic balance condition and P. Syty for remarks on the manuscript.
The work is supported by the Committee for Scientific Research (Poland)
under project No. 2P03B 000912.
The support from Foundation for Polish Science is also gratefully acknowledged.

\begin{table}
\caption[Coefficients]
{\label{wsp} Elements of expansions of sine- and cosine-like solutions
in the Laguerre and the Gaussian basis sets \cite{matem}. The $L_{n}^{(\alpha)}$
and $C_{n}^{(\alpha)}$ are the Laguerre and the Gegenbauer polynomials,
respectively  while ${}_{2}F_{1}$ and ${}_{1}F_{1}$ are the
Gauss and the Kummer (confluent)
hypergeometric functions (see \cite{erd2}) respectively; $\lambda > 0 $ is a
scaling parameter.}
\begin{center}
\begin{tabular}{ccc}
Quantity & Laguerre set & Gaussian set \\
\hline
$\phi_{n}^{l}$ &
$(\lambda r)^{l+1}\exp(- \lambda r/2)\:L_{n}^{(2l+1)}(\lambda r)$ &
$(\lambda r)^{l+1}\exp(- \lambda^{2} r^{2}/2)\:L_{n}^{(2l+1)}(\lambda^{2} r^{2})$ \\
$\bar{\phi}_{n}^{l}$ &
$\frac{n!}{\lambda n+2l+1}(\lambda r)^{-1}\phi_{n}^{l} $ &
$\frac{2n!}{\lambda^{2} \Gamma(n+2l+3/2)}\phi_{n}^{l}$ \\
$s^{l}_{n}$&
${ 2^l l! n ! (\sin\theta)^{l+1}
  \over n+2l+1}C_{n}^{(l+1)}(\cos\theta)$&
${\sqrt{2\pi} n ! (-1)^{n}
 \over \Gamma(n+l+\frac{3}{2})}\exp(- \eta^{2}/2)L^{(l+1/2)}_{n}(\eta^2)$ \\
$c^{l}_{n}$ & ${- 2^l\Gamma(l+{1 \over 2}) n !
  \over \sqrt{\pi}\Gamma(n+2l+2)(\sin\theta)^{l} } $&
${\sqrt{2 \over \pi}\Gamma(n+\frac{1}{2})   (-1)^{n} n !
\over \Gamma(n+l+\frac{3}{2})}\exp(- \eta^{2}/2) {\eta}^{-l} $ \\
&$\times {}_{2}F_{1}(-n-2l-1,n+1;\frac{1}{2}-l; \sin^{2}(\theta/2))$,&
$\times {}_{1}F_{1}(-n-l-\frac{1}{2},\frac{1}{2}-l,\eta^{2})$, \\
&$\sin\theta\equiv\frac{k \lambda^{-1}}{k^2 \lambda^{-2} + {1 \over 4}}$&
$\eta\equiv\frac{k}{\lambda}$\\
\end{tabular}
\end{center}
\end{table}
\begin{table}
\caption[Matrix elements]
{\label{relmatrix} Biorthonormal elements $\bar{\psi}_{n}^{l}$ due
to the small component.}
\begin{center}
\begin{tabular}{ccc}
Quantity & Laguerre set & Gaussian set \\
\hline
$\bar{\psi}_{n}^{l}$ for $\kappa>0$&
$-\frac{n!}{n+2l+1}(\lambda r)^l\exp(-\lambda r/2)
 $
&$-\frac{n!}{\lambda^{2} \Gamma(n+2l+3/2)}\exp(-\lambda^{2} r^{2}/2) (\lambda r)^l
 $\\
& $\times\sum_{k=0}^n(-2)^{k+1} L^{(2l+k+1)}_{n-k}(\lambda r)$ &
$\times \sum_{k=0}^n(-2)^{k+1} L^{(2l+k+1)}_{n-k}(\lambda^{2}r^{2})$ \\
$\bar{\psi}_{n}^{l}$ for $\kappa <0$ &$-\frac{n!}{n+2l+1}\exp(-\lambda r/2)
\sum_{k=0}^n(-2)^{k+1} $ &
$-\frac{n!}{\lambda^{2} \Gamma(n+2l+3/2)} (\lambda r)^{-(l+1)} $
\\
&$\times \sum_{i=0}^{k}\frac{(2l+1)!}{(2l+1-i)!} (\lambda r)^{2l+1-i}(-1)^{i} L^{(2l+1+i)}_{n-i}(\lambda r)$&
$\times\int_{0}^{(\lambda r)^{2}}t^{l+1/2}\exp(-t^{2}/2)L_{n}^{(l+1/2)}(t)dt$ \\
\end{tabular}
\end{center}
\end{table}

\begin{table}
\caption[Matrix elements 1]
{\label{relmatrix1}The overlap integrals proportional to the elements
of ${\cal J}$-matrix.}
\begin{center}
\begin{tabular}{ccc}
Integral & Laguerre set & Gaussian set \\
\hline
$\langle \phi_{m}^{l} | \phi_{n}^{l} \rangle $ &
 $\frac{\Gamma(n+2l+2)}{\lambda n! }[2(n+2l+2)\delta_{mn}-n\delta_{m,n-1}$
&$\frac{\Gamma(n+l+3/2)}{2 n!} \delta_{mn}$ \\
&$-(n+2n+3)\delta_{m,n+1}]$ & \\
$\langle \psi_{m}^{l} | \psi_{n}^{l} \rangle $ &
$\frac{\Gamma(n+2l+2)}{4 n! }[2(n+2l+2)(2 \lambda -1)\delta_{mn}+$
& $\frac{\lambda^{2} \Gamma(n+l+ 3/2)}{2 n!}[(2n+l+3/2)
\delta_{mn}+ $ \\
&$n\delta_{m,n-1}+(n+2n+3)\delta_{m,n+1}]$ & $n\delta_{m,n-1}+(n+l+3/2)\delta_{m,n+1}]$ \\
\end{tabular}
\end{center}
\end{table}


\begin{thebibliography}{99}
\bibitem{Hell74a}
E. Heller and H. Yamani, Phys. Rev. A {\bf 9} (1974) 1201.
\bibitem{Hell74b}
E. Heller and H. Yamani, Phys. Rev. A {\bf 9} (1974) 1209.
\bibitem{matem}
H. Yamani and L. Fishman, J. Math. Phys.  {\bf 16} (1979) 410.
\bibitem{Hell}
E. J. Heller, Phys. Rev. A, {\bf 12} (1975) 1222.
\bibitem{Broad}
J. T. Broad, Phys. Rev. A, {\bf 18} (1978) 1012.
\bibitem{YR}
H. Yamani and W. P. Reinhardt,  Phys. Rev. A {\bf 11} (1975) 1144.
\bibitem{Yamani}
H. A. Yamani, J. Math. Phys. {\bf 25} (1984) 317.
\bibitem{YA1}
H. Yamani and M. S. Abdelmonem, J. Phys. A: Math. Gen.
{\bf 29} (1996) 6991.
\bibitem{YA2}
H. Yamani and M. S. Abdelmonem, J. Phys. B: At. Mol. Opt. Phys.
{\bf 30} (1997) 1633.
\bibitem{Harris}
F. E. Harris, Phys. Rev. Lett. {\bf 19} (1967) 173.
\bibitem{erd2}
A. Erd\'{e}lyi, Ed., Higher Transcendental Functions, Vol. II
(McGraw-Hill, New York, 1953).
\bibitem{GoldPRA}
S. P. Goldman, Phys. Rev. A {\bf 40} (1998) 1185.
\bibitem{Dyall1}
K. G. Dyall, I. P. Grant and S. Wilson, J. Phys. B: At. Mol. Phys.
{\bf 17} (1984) L45.
\bibitem{Dyall2}
K. G. Dyall, I. P. Grant and S. Wilson, J. Phys. B: At. Mol. Phys.
{\bf 17} (1984) 1201.
\bibitem{Grant}
I. P. Grant, J. Phys. B: At. Mol. Phys. {\bf 19} (1986) 3187.
\bibitem{Schw1}
W. H. E. Schwarz and H. Wallmeier, Mol. Phys. {\bf 46} (1982) 1045.
\bibitem{Schw2}
W. H. E. Schwarz and E. Wechsel-Tarkowski,
Chem. Phys. Lett. {\bf 85}, (1982) 94.
\bibitem{Kutz}
W. Kutzelnigg, Int. J. Quantum Chem. {\bf 25} (1984) 107.
\bibitem{Radek}
R. Szmytkowski, J. Phys. B: At. Mol. Phys. {\bf 30} (1997) 825;
J. Phys. B: At. Mol. Phys. {\bf 30} (1997) 2747.
\bibitem{Gold}
G. W. F. Drake and S. P. Goldman, Adv. Atom. Mol. Phys. {\bf 25}
(1988) 393.
\end{thebibliography}
\end{document}